\documentclass{bmcart}

\usepackage{amsthm,amsmath}
\usepackage{bm}
\usepackage{url}
\usepackage{dsfont}
\usepackage{graphicx}
\usepackage{multirow}
\usepackage[utf8]{inputenc} 

\startlocaldefs
\endlocaldefs

\DeclareMathOperator{\logit}{logit}

\begin{document}

\begin{frontmatter}

\begin{fmbox}
\dochead{Research}


\title{Model-based standardization using multiple imputation}


\author[
   addressref={aff1},                   
   corref={aff1},                       
   email={antonio.remiro-azocar@bayer.com}   
]{\inits{ARA}\fnm{Antonio} \snm{Remiro-Az\'ocar}}
\author[
   addressref={aff2,aff3,aff4},                   
   email={anna.heath@sickkids.ca}   
]{\inits{AH}\fnm{Anna} \snm{Heath}}
\author[
   addressref={aff4},                   
   email={g.baio@ucl.ac.uk}   
]{\inits{GB}\fnm{Gianluca} \snm{Baio}}


\address[id=aff1]{
  \orgname{Statistics and Data Insights, Bayer plc}, 
  \street{400 South Oak Way},                     %
  \city{Reading},                              
  \cny{UK}                                    
}

\address[id=aff2]{%
  \orgname{Child Health Evaluative Sciences, The Hospital for Sick Children},
  \street{686 Bay Street},  
  \city{Toronto},
  \cny{Canada}
}

\address[id=aff3]{%
  \orgname{Dalla Lana School of Public Health, University of Toronto},
  \street{115 College Street},
  \city{Toronto},
  \cny{Canada}
}

\address[id=aff4]{%
  \orgname{Department of Statistical Science, University College London},
  \street{1-19 Torrington Place},
  \city{London},
  \cny{UK}
}


\begin{artnotes}
\note{$^1$Statistics and Data Insights, Bayer plc, Reading, UK}
\note{$^2$Child Health Evaluative Sciences, Hospital for Sick Children, Toronto, Canada}
\note{$^3$Dalla Lana School of Public Health, University of Toronto, Canada}
\note{$^4$Department of Statistical Science, University College London, UK}            
\end{artnotes}

\end{fmbox}


\begin{abstractbox}

\begin{abstract} 
\textbf{Background:} When studying the association between treatment and a clinical outcome, a parametric multivariable model of the conditional outcome expectation is often used to adjust for covariates. The treatment coefficient of the outcome model targets a conditional treatment effect. Model-based standardization is typically applied to average the model predictions over the target covariate distribution, and generate a covariate-adjusted estimate of the marginal treatment effect. 
\textbf{Methods:} The standard approach to model-based standardization involves maximum-likelihood estimation and use of the non-parametric bootstrap. We introduce a novel, general-purpose, model-based standardization method based on multiple imputation that is easily applicable when the outcome model is a generalized linear model. We term our proposed approach multiple imputation marginalization (MIM). MIM consists of two main stages: the generation of synthetic datasets and their analysis. MIM accommodates a Bayesian statistical framework, which naturally allows for the principled propagation of uncertainty, integrates the analysis into a probabilistic framework, and allows for the incorporation of prior evidence.
\textbf{Results:} We conduct a simulation study to benchmark the finite-sample performance of MIM in conjunction with a parametric outcome model. The simulations provide proof-of-principle in scenarios with binary outcomes, continuous-valued covariates, a logistic outcome model and the marginal log odds ratio as the target effect measure. When parametric modeling assumptions hold, MIM yields unbiased estimation in the target covariate distribution, valid coverage rates, and similar precision and efficiency than the standard approach to model-based standardization. \textbf{Conclusion:} We demonstrate that multiple imputation can be used to marginalize over a target covariate distribution, providing appropriate inference with a correctly specified parametric outcome model and offering statistical performance comparable to that of the standard approach to model-based standardization. 
\end{abstract}


\begin{keyword}
\kwd{standardization}
\kwd{marginalization}
\kwd{multiple imputation}
\kwd{parametric G-computation}
\kwd{covariate adjustment}
\kwd{indirect treatment comparisons}
\end{keyword}


\end{abstractbox}
%

\end{frontmatter}




\section*{Background}

There has been active debate on whether marginal or conditional estimands should be preferred when estimating relative treatment effects \cite{remiro2022target, russek2022discussion, spieker2022comments, senn2022conditions, schiel2022commentary, van2022estimands}. Many researchers argue that marginal estimands are more appropriate inferential targets for decisions at the population level \cite{remiro2022target, russek2022discussion, spieker2022comments}. The distinction between marginal and conditional treatment effects is particularly important for non-collapsible measures such as the odds ratio and the hazard ratio. For such measures, the population-level marginal effect cannot be expressed as a weighted average of individual- or subgroup-level conditional effects. Almost invariably, marginal and conditional estimands do not coincide for non-collapsible effect measures, even in the absence of confounding and effect measure modification \cite{greenland2011adjustments, greenland2001confounding, kaufman2010marginalia, whittemore1978collapsibility, greenland1999confounding, huitfeldt2019collapsibility}.

In the estimation of marginal treatment effects, covariate adjustment is desirable across a range of settings: (1) it is applied in the analysis of randomized controlled trials (RCTs) to correct for ``chance'' covariate imbalances and increase power, precision and efficiency \cite{morris2022planning}; (2) it allows for confounding control in the analysis of observational studies \cite{austin2013performance}; and (3) it accounts for covariate differences between multiple studies in indirect treatment comparisons and transportability analyses \cite{remiro2021methods, remiro2022parametric, remiro2022two, josey2021transporting, phillippo2020multilevel}. This article focuses on the latter scenario. Nevertheless, the methodological findings are also applicable to covariate adjustment between the treatment arms of a single comparative study. 

A popular approach to covariate adjustment involves fitting a parametric multivariable model of the conditional outcome mean given treatment and baseline covariates. The treatment coefficient of the model targets a conditional effect. So-called model-based standardization, G-computation or marginalization approaches are required to integrate or average the conditional outcome model over the target covariate distribution, and produce a covariate-adjusted estimate of the marginal treatment effect \cite{remiro2022parametric, robins1986new, zhang2008estimating, moore2009covariate, austin2010absolute, rosenblum2010simple, snowden2011implementation, wang2017g, daniel2021making, campbell2023standardization, vo2019novel}. The standard approach to model-based standardization uses maximum-likelihood estimation to fit the outcome model and the non-parametric bootstrap for variance estimation \cite{remiro2022parametric, snowden2011implementation, wang2017g, daniel2021making, campbell2023standardization}.

We introduce a novel general-purpose method for model-based standardization stemming from the ideas underlying multiple imputation \cite{rubin2004multiple}. Despite the close relationships between the methodologies, these largely have been developed separately, with some exceptions \cite{westreich2015imputation}. We build a link in this article,\footnote{This article is based on research from Antonio Remiro-Az\'ocar's PhD thesis \cite{remiro2022population} and a prior working paper by the authors \cite{remiro2020marginalization}.} terming our proposed approach \textit{multiple imputation marginalization} (MIM). As opposed to the standard version of model-based standardization, MIM accommodates a Bayesian statistical framework, which naturally allows for the principled propagation of uncertainty, readily handles missingness in the patient-level data, integrates the analysis into a probabilistic framework, and permits the incorporation of prior evidence and other contextual information. 

We conduct a simulation study to benchmark the finite-sample performance of MIM in conjunction with a parametric outcome model. The simulations provide proof-of-principle in scenarios with binary outcomes, continuous-valued covariates, a logistic outcome model and the marginal log odds ratio as the target effect measure. When parametric modeling assumptions hold, MIM yields unbiased estimation in the target covariate distribution. Code to implement the MIM methodology in \texttt{R} is provided in Additional file 1.
  
\section*{Methods}

We wish to transport the results of a comparative ``index'' study to a target distribution of covariates. We assume that the target is characterized by a dataset that is external to the index study. In practice, this could belong to an observational study or be derived from secondary healthcare data sources (e.g.~disease registries, cohort studies, insurance claims databases or electronic health records). Such administrative datasets are typically larger, less selected, and more representative of target populations of policy-interest than the participants recruited by RCTs \cite{girman2019considerations, weiss2019generalizing, ramsey2020using}. 

For instance, in drug development, a pivotal Phase III RCT is typically conducted pre-market authorization to obtain regulatory approval. Such trial may have relatively narrow selection criteria, to enhance statistical precision and power in efficacy and safety testing \cite{rothwell2005external, rothwell2010commentary, greenhouse2008generalizing}. Policy-makers may be interested in transporting inferences to a ``real-world'' target covariate distribution, which is more diverse or heterogeneous in composition, and more representative of the patients who will receive the intervention in routine clinical practice \cite{happich2020reweighting}. 

Let $S=1$ denote the index study and let $S=2$ denote the external target. Adopting the potential outcomes framework \cite{rubin1974estimating}, the target marginal average treatment effect estimand for the MIM procedure described in this article is a contrast between the, possibly transformed, means of potential outcome distributions:
\begin{equation}
TATE = g\left(E\left(Y^1 \mid S=2 \right )\right) - g\left(E\left(Y^0 \mid S=2 \right)\right),
\label{tate_additive}
\end{equation}
where $Y^t$ denotes the potential outcome that would have been observed for a subject assigned to intervention $T=t$, with $t \in \{0, 1\}$, $E(\cdot)$ represents an expectation taken over the distribution of potential outcomes in $S=2$, and $g(\cdot)$ is an appropriate ``link'' function, e.g.~the identity, log or logit, mapping the mean potential outcomes onto the plus/minus infinity range. The target estimand in Equation \ref{tate_additive} is the average treatment effect, constructed on an additive scale e.g.~the mean difference, (log) risk ratio or (log) odds ratio scale, had everyone in the target had been assigned $T=1$ versus $T=0$. 
 
We briefly outline the data requirements of the MIM procedure. Individual-level data $\mathcal{D}=\left(\bm{x}, \bm{t}, \bm{y} \right)$ for a comparative index study, randomized or non-randomized, are available. Here, $\bm{x}$ is an $N \times K$ matrix of clinical or demographic baseline covariates, where $N$ is the number of participants in the study and $K$ is the number of baseline covariates. Each subject $n=1,2,\dots,N$ has a row vector $\bm{x}_n = \left(x_{n,1}, x_{n,2}, \dots, x_{n,K}\right)$ of $K$ covariates. We let $\bm{y} = \left(y_1, y_2, \dots, y_N\right)$ denote a vector of clinical outcomes and $\bm{t}=\left(t_1, t_2, \dots, t_N\right)$ denote a binary treatment indicator vector, with each entry taking the value zero or one. We assume that $\mathcal{D}$ has no missing values but MIM can be readily adapted to address this issue, as is illustrated in Additional file 1. 

The target dataset contains a matrix of covariates $\bm{x}^{tar}$ of dimensions $N^{tar} \times K$, where $N^{tar}$ is the number of subjects and $K$ is the number of covariates. We assume that all $K$ covariates in the index study are available in the target. Each subject has a row vector $\bm{x}_i^{tar}=\left(x_{i,1}^{tar}, x_{i,2}^{tar}, \dots, x_{i,K}^{tar}\right)$ of $K$ covariates. Individual-level outcomes under the treatments being studied in the index trial are assumed unavailable in the target, as would be the case for interventions evaluated in the pre-marketing authorization setting.

In the scenario described in this article, standardization is performed with respect to an external data source, and the aim is to estimate marginal treatment effects in an external covariate distribution. This is typically the case in transportability analyses translating inferences from trials lacking external validity to the target population for decision-making, or in covariate-adjusted indirect comparisons transporting relative effects across a connected network of trials. 

Nevertheless, as illustrated in Additional file 1, it is also possible to perform standardization over the covariate distribution observed in the index study. This avoids extrapolation into an external data source and may be useful when adjusting for covariate imbalances between treatment arms within randomized or non-randomized comparative studies. Within a randomized experiment, covariate adjustment is not necessary for unbiased estimation of the marginal treatment effect, but can be used to increase precision, i.e.~reduce standard errors \cite{morris2022planning, tackney2023comparison}. Within a non-randomized study, covariate adjustment is necessary to remove confounding bias \cite{austin2011introduction}. 

\subsection*{Multiple imputation marginalization}

Conceptually, MIM consists of two separate stages: (1) the generation (\textit{synthesis}) of synthetic datasets; and (2) the \textit{analysis} of the generated datasets. The synthesis is separated from the analysis --- only after the synthesis has been completed is the marginal effect of treatment on the outcome estimated. This is analogous to the separation between the imputation and analysis stages in multiple imputation. 

Multiple imputation is a simulation technique that, arguably, is fundamentally Bayesian \cite{rubin2004multiple, meng1994multiple, gabrio2019full}. Its original development was grounded in Bayesian modeling, with imputed outcomes derived, at least conceptually, from a posterior predictive distribution. Computational tools such as Markov chain Monte Carlo (MCMC) and the Gibbs sampler only arose to prominence in the statistical literature several years after Rubin's seminal paper \cite{rubin1978multiple}. Consequently, the typical practical implementation of multiple imputation is based on a hybrid approach \cite{rubin2004multiple}: ``think like a Bayesian, do like a frequentist''.

Interestingly, our standardization problem can be conceptualized as a missing data problem. Outcomes for the subjects in the index study are observed, but outcomes in the target population, under the treatments examined in the index study, are systematically missing. MIM standardizes over the target by replacing the missing outcomes with a set of plausible values, conditional on some pre-specified imputation mechanism. Extending the parallel with the missing data literature, MIM relies on a missing-at-random-like assumption: missing outcomes in the target are assumed conditionally exchangeable with those observed in the index study, conditioning on the adjustment model used for standardization. 

MIM sits within a Bayesian framework by characterizing probabilistic relationships among a set of variables, and adopts a simulation approach. Figure \ref{fig1} reveals a Bayesian directed acyclic graph (DAG) summarizing the general MIM structure and the links between its modules. In this graphical representation, the nodes represent variables; single arrows indicate probabilistic relationships and double arrows indicate deterministic functions. The plate notation indicates repeated analyses. We return to Figure \ref{fig1} and provide more detailed explanations for the notation and the individual modules throughout this section. 

\subsection*{Generation of synthetic datasets: a missing data problem}

The first stage, synthetic data generation, consists of two steps. Initially, the \textit{first-stage regression} captures the relationship between the outcome $\bm{y}$ and the covariates $\bm{x}$ and treatment $\bm{t}$ in the patient-level data for the index study. In the \textit{outcome prediction} step, predicted outcomes for each treatment are generated in the target by drawing from the posterior predictive distribution of outcomes, given the observed predictor-outcome relationships in the index study, the set treatment and the target covariate distribution. 

\subsubsection*{First-stage regression}

Firstly, a multivariable regression of the observed outcome $\bm{y}$ on the baseline covariates $\bm{x}$ and treatment $\bm{t}$ is fitted to the subject-level data of the index study: 
\begin{equation}
g\left(\mu_n\right) = \beta_0 +  \bm{x}_n\bm{\beta_1} + \left(\beta_t +  \bm{x}_n \bm{\beta_2}\right)1\left(t_n=1\right),
\label{equation8}
\end{equation}
where $\mu_n$ is the conditional outcome expectation of subject $n$ on the natural scale (e.g.,~the probability scale for binary outcomes), $g\left(\cdot\right)$ is an appropriate link function, $\beta_0$ is the intercept, $\bm{\beta_1}$ and $\bm{\beta_2}$ are vectors of regression coefficients, and the treatment coefficient $\beta_t$ targets a conditional effect at baseline, when the covariates are zero. The model specification assumes that the covariates are prognostic of outcome at the individual level. Due to the presence of treatment-covariate interactions, the covariates are also assumed to be (conditional) effect measure modifiers, i.e.,~predictive of treatment effect heterogeneity, at the individual level on the linear predictor scale. 

The conditional outcome model in Equation \ref{equation8} will be our ``working'', ``nuisance'' or ``imputation'' model from now onward. We consider this to be a parametric model within a generalized linear modeling framework. In logistic regression, the link function $g\left(\mu_n\right)=\logit\left(\mu_n\right) = \ln \left[\mu_n/\left(1-\mu_n\right)\right]$ is adopted. Other choices are possible in practice such as the identity link for linear regression or the log link for Poisson regression. The conditional outcome model is to be estimated using a Bayesian approach. We shall assume that efficient simulation-based sampling methods such as MCMC are used. Prior distributions for the regression coefficients would have to be specified, potentially using contextual information. 

When standardizing with respect to an external target and/or when the index study is non-randomized, one is reliant on correct specification of the outcome model for unbiased estimation. In the former case, there is particular interest in modeling covariate-treatment product terms (``interactions'') to capture (conditional) effect measure modification. In the latter case, the outcome model should adjust for potential confounders. Time and care should be dedicated to model-building, while being mindful of erroneous extrapolation outside the covariate space observed in the index study \cite{vo2023cautionary}.  

We shall assume that a single parametric outcome model is estimated, including treatment-covariate product terms to capture treatment effect heterogeneity at the individual level. An alternative strategy is to postulate two separate outcome models, one for each treatment group in the index comparative study \cite{dahabreh2020extending}. While such approach allows for individual-level treatment effect heterogeneity over all the baseline covariates included in the models, it prevents borrowing information across treatment groups. 

\subsubsection*{Outcome prediction}

In this step, we generate predicted outcomes for the treatments under investigation, but in the target covariate distribution, by drawing from the posterior predictive distribution of outcomes. This is to be constructed using the imputation model in Equation \ref{equation8}. Beforehand, a ``data augmentation'' step is required. We shall create a copy of the original target covariate dataset and vertically concatenate it to the original $\bm{x}^{tar}$. The concatenation is denoted $\bm{x^*}=\left[ \bm{x}^{tar} \atop \bm{x}^{tar} \right]$ and has $N^*=\left(2 \times N^{tar}\right)$ rows and $K$ columns. The original $j=1, 2, \dots, N^{tar}$ rows are assigned the treatment value $t^{*}_j=1$ and the appended $j=\left(N^{tar}+1\right), \left(N^{tar}+2\right) \dots, N^{*}$ rows are assigned the treatment value $t^{*}_j=0$. The treatment indicator vector in the augmented dataset is denoted $\bm{t^*}=\left(t^*_1, t^*_2 \dots t^*_{N^*}\right)$. 

Using MCMC sampling, it is fairly straightforward to implement the estimation of both the first-stage regression and the outcome prediction steps within a single Bayesian computation module. Having fitted the first-stage regression, we will iterate over the $L$ converged draws of the MCMC algorithm to generate $M \leq L$ synthetic datasets: $\{ \mathcal{D}^{*} = \mathcal{D}^{*(m)}: m=1,2\dots,M \}$, where $\mathcal{D}^{*(m)}=\left(\bm{x}^{\bm{*}}, \bm{t^*}, \bm{y}^{\bm{*}(m)}\right)$. Covariates $\bm{x^*}$ and treatment $\bm{t^*}$ are fixed across all the synthetic datasets. In line with the multiple imputation framework, each synthetic dataset is filled in by drawing a vector of outcomes $\bm{y}^{\bm{*}(m)} = \left(y^{*(m)}_1, y^{*(m)}_2, \dots, y^{*(m)}_{N^*}\right)$ of size $N^*$ from the posterior predictive distribution $p\left(\bm{y^*} \mid \bm{x^*}, \bm{t^*}, \bm{y}, \bm{x}, \bm{t}\right)$, given the original index trial and the augmented target datasets. 

Assuming convergence of the MCMC sampling algorithm, the posterior predictive distribution of outcomes is approximated numerically as: \begin{align*}
p\left(\bm{y^*} \mid \bm{x^*}, \bm{t^*}, \bm{y}, \bm{x}, \bm{t}\right) &= \int_{\bm{\beta}} p \left(\bm{y^*} \mid \bm{x^*}, \bm{t^*}, \bm{\beta} \right) \cdot p\left(\bm{\beta} \mid \bm{y}, \bm{x}, \bm{t}\right) d \bm{\beta} \\
&\approx
\frac{1}{L}
\sum_{l=1}^L p\left(\bm{y^*} \mid \bm{x^*}, \bm{t^*}, \bm{\beta}^{(l)}\right),
\end{align*}
where the realizations $\bm{\beta}^{(l)} \sim p \left(\bm{\beta} \mid \bm{y}, \bm{x}, \bm{t}\right)$ are independent draws from the posterior distribution of the first-stage regression parameters $\bm{\beta} = \left(\beta_0, \bm{\beta_1}, \bm{\beta_2}, \beta_t \right)$, which encode the predictor-outcome relationships observed in the index trial, given some suitably defined prior $p\left(\bm{\beta}\right)$. Here, $l=1, 2, \dots L$ indexes each MCMC iteration after convergence. 

Consequently, $L$ predictive samples $\bm{y}^{\bm{*}(1)}, \dots, \bm{y}^{\bm{*}(L)} \sim p\left(\bm{y^*} \mid \bm{x^*}, \bm{t^*}, \bm{\beta}^{(l)}\right)$ are drawn independently from the posterior predictive distribution of outcomes. Dedicated MCMC programming software such as \texttt{Stan} \cite{carpenter2017stan} would typically return an $L \times N^*$ matrix of simulations. We will ``thin'' the matrix such that only $M \leq L$ of the rows are retained. This is to reduce the computational run time of the MIM analysis stage. The $M$ remaining outcome imputations are used to complete the synthetic datasets $\{\mathcal{D}^{*(m)}=\left(\bm{x}^{\bm{*}}, \bm{t^*}, \bm{y}^{\bm{*}(m)}\right): m=1,2,\dots,M\}$. Table \ref{tab1} illustrates the structure of each synthetic dataset. 

\subsection*{Analysis of synthetic datasets}

In the second stage, the analysis of synthetic datasets, we seek inferences about the marginal treatment effect in the target covariate distribution ($TATE$ in Equation \ref{tate_additive}, but here denoted $\Delta$), given the synthesized outcomes. The analysis stage consists of another two steps. In the \textit{second-stage regression} step, estimates of the marginal treatment effect in each synthesis $m=1,2,\dots,M$ are generated by regressing the predicted outcomes $\bm{y}^{\bm{*}(m)}$ on the treatment indicator $\bm{t^{*}}$. In the \textit{pooling} step, the treatment effect estimates and their variances are combined across all $M$ syntheses.

In standard multiple imputation, the imputation and analysis stages may be performed simultaneously in a joint model \cite{gabrio2019full}. In MIM, this is challenging because the dependent variable of the analysis is completely synthesized. Consider the Bayesian DAG in Figure \ref{fig1}. In a joint model, the predicted outcomes are a collider variable, blocking the only path between the first and the second module (information from the directed arrows ``collides'' at the node). As a result, the data synthesis and analysis stages have been implemented as separate modules in a two-stage framework. The analysis stage conditions on the outcomes predicted by the synthesis stage, treating these as observed data. 

\subsubsection*{Second-stage regression}

We fit $M$ second-stage regressions of predicted outcomes $\bm{y}^{\bm{*}(m)}$ on treatment $\bm{t^*}$ for $m=1,2,\dots,M$. Identical analyses are performed on each synthesis:
\begin{equation}
g\left(\eta_j^{(m)}\right) = \alpha^{(m)} + \delta^{(m)} t^*_j,
\label{equation24}
\end{equation}
where $\eta_j^{(m)}$ is the expected outcome on the natural scale of unit $j$ in the $m$-th synthesis, the coefficient $\alpha^{(m)}$ is an intercept term and $\delta^{(m)}$ denotes the marginal treatment effect in the $m$-th synthesis. There is some non-trivial computational complexity to performing a Bayesian fit in this step. That would embed a nested simulation scheme. Namely, if we draw $M$ samples 
$\{ \bm{y}^{\bm{*}(m)} : m = 1,2,\dots M \}$ in the synthesis stage, a further number of samples, say $R$, of the treatment effect $\{ \delta^{(m,r)}: m = 1,2\dots M; r = 1, 2,\dots R \}$ would be drawn for each of these realizations separately. This structure is unlikely to be feasible in terms of running time. 

Using maximum-likelihood estimation, a point estimate $\hat{\delta}^{(m)}$ of the marginal treatment effect and a measure of its variance $\hat{v}^{(m)}$ are generated in each synthesis $\bm{y}^{\bm{*}(m)}$. Equation \ref{equation24} is a marginal model of outcome on treatment alone. Adopting terminology from the missing data literature, the second-stage regression in the analysis stage is ``congenial'' with the first-stage regression in the synthesis stage because treatment was already included as a predictor in the first-stage regression \cite{meng1994multiple}.

\subsubsection*{Pooling}

We must now combine the $M$ point estimates of the marginal treatment effect and their variances to generate a posterior distribution. Pooling across multiple syntheses is a topic that has already been investigated within the domain of statistical disclosure limitation \cite{raghunathan2003multiple, rubin1993statistical, reiter2002satisfying, reiter2005releasing, si2011comparison, reiter2007multiple, raab2016practical}.

In statistical disclosure limitation, data agencies mitigate the risk of identity disclosure by releasing multiple \textit{fully synthetic} datasets. These only contain simulated values, in lieu of the original confidential data of real survey respondents. Raghunathan et al. \cite{raghunathan2003multiple} describe full synthesis as a two-step process: (1) construct multiple synthetic populations by repeatedly drawing from the posterior predictive distribution, conditional on a model fitted to the original data; and (2) draw random samples from each synthetic population, releasing these synthetic samples to the public. In practice, as indicated by Reiter and Raghunathan \cite{reiter2007multiple}, it is not a requirement to generate the populations, but only to generate values for the synthetic samples. Once the samples are released, the analyst seeks inferences based on the synthetic data alone.  

MIM is analogous to this problem, albeit there are some differences. In MIM, the analyst also acts as the synthesizer of data, and there is no ``original data'' on outcomes as such if the index study has not been conducted in the target covariate distribution. In any case, values for the samples are generated in the synthesis stage by repeatedly drawing from the posterior predictive distribution of outcomes. This is conditional on the predictor-outcome relationships indexed by the model fitted to the index study, the set treatment and the target distribution of covariates. 

We seek to construct a posterior distribution for the marginal treatment effect, conditional on the synthetic outcomes (and treatment). That is, $p\left(\Delta \mid \bm{y}^{\bm{*}}, \bm{t^*}\right)$. Following Raab et al. \cite{raab2016practical}, each $\bm{y}^{\bm{*}(m)}$ is viewed as a random sample from $p\left(\bm{y}^{\bm{*}} \mid \bm{x}^{\bm{*}}, \bm{t^*}, \bm{\beta}^{(m)} \right)$, where $\bm{\beta}^{(m)}$ is sampled from its posterior $p\left(\bm{\beta} \mid \bm{x}, \bm{t}, \bm{y} \right)$. Hence, the ``true'' marginal treatment effect $\delta^{(m)}$ for the $m$-th synthesis, corresponding to $\bm{\beta}^{(m)}$, can be defined as a function of this sample. In each second-stage regression in Equation \ref{equation24}, this is the treatment effect estimated by $\hat{\delta}^{(m)}$. 

Consequently, following Raghunathan et al. \cite{raghunathan2003multiple}, the estimators $\{ \hat{\delta}^{(m)}, \hat{v}^{(m)}; m=1,2,\dots,M \}$ from the second-stage regressions are treated as ``data'', and are used to construct an approximation to the posterior density $p\left(\Delta \mid \bm{y}^{\bm{*}}, \bm{t^*}\right)$. This density is assumed to be approximately normal and is parametrized by its first two moments: the mean $\mu_{\Delta}$, and the variance $\sigma_{\Delta}^2$. To derive the conditional distribution $p\left(\mu_\Delta, \sigma_\Delta^2 \mid
\bm{y}^{\bm{*}}, \bm{t^*}\right)$ of these moments given the syntheses, the estimators $\{ \hat{\delta}^{(m)}, \hat{v}^{(m)}; m=1,2,\dots,M \}$, where $\hat{v}^{(m)}$ is the point estimate of the variance in the $m$-th second-stage regression, are treated as sufficient summaries of the syntheses, and $\mu_{\Delta}$ and $\sigma_{\Delta}^2$ are treated as parameters. Then, the posterior distribution $p\left(\Delta \mid \bm{y}^{\bm{*}}, \bm{t^*}\right)$ is constructed as:
\begin{equation}
p\left(\Delta \mid \bm{y}^{\bm{*}}, \bm{t^*}\right) = \int_{\mu_\Delta, \sigma_\Delta^2} p\left(\Delta
\mid \mu_\Delta, \sigma_\Delta^2 \right)
p\left(\mu_\Delta, \sigma_\Delta^2 \mid 
\bm{y}^{\bm{*}}, \bm{t^*}\right)
d\left(\mu_\Delta, \sigma_\Delta^2\right).
\label{equation25}
\end{equation}

In analogy with the theory of multiple imputation \cite{rubin2004multiple}, the following quantities are required for inference: 
\begin{align}
\bar{\delta} &= \sum_{m=1}^M
\hat{\delta}^{(m)}/M, 
\label{equation26}
\\
\bar{v} &= \sum_{m=1}^M \hat{v}^{(m)}/M, 
\label{equation27}
\\
b &= \sum_{m=1}^{M}\left (\hat{\delta}^{(m)} - \bar{\delta}\right)^2/\left(M-1\right),
\label{equation28}
\end{align}
where $\bar{\delta}$ is the average of the treatment effect point estimates across the $M$ syntheses, $\bar{v}$ is the average of the point estimates of the variance (the ``within'' variance), and $b$ is the sample variance of the point estimates (the ``between'' variance). These quantities are computed using the point estimates from the second-stage regressions. 

After deriving the quantities in Equations \ref{equation26}, \ref{equation27} and \ref{equation28}, there are two options to approximate the posterior distribution of the marginal treatment effect in Equation \ref{equation25}. The first involves direct Monte Carlo simulation and the second uses a simple normal approximation. In Additional file 1, the inferential framework for pooling outlined in this section is extended to scenarios involving correlated outcomes and non-scalar estimands with multiple components \cite{bujkiewicz2019multivariate}. This involves a multivariate outcome model (i.e.~with multiple dependent variables) and the combination of correlated treatment effects corresponding to multiple outcomes. 

\paragraph{Pooling via posterior simulation}

Firstly, one draws $\mu_\Delta$ and $\sigma_\Delta^2$ from their posterior distributions, conditional on the syntheses. These distributions are derived by Raghunathan et al. \cite{raghunathan2003multiple}. Values of $\mu_\Delta$ are drawn from a normal distribution:
\begin{equation}
p\left(\mu_\Delta \mid \bm{y}^{\bm{*}}, \bm{t^*}\right) \sim \textnormal{N}\left(\bar{\delta}, \bar{v}/M \right),
\label{equation29}
\end{equation}
Values of $\sigma_\Delta^2$ are drawn from a chi-squared distribution with $M-1$ degrees of freedom:
\begin{equation}
p\left(\left(M-1\right)b/\left(\sigma_\Delta^2 + \bar{v}\right) \mid \bm{y}^{\bm{*}}, \bm{t^*}\right)
\sim \chi^2_{M-1}.
\label{equation30}
\end{equation}
Values of $\Delta$ are drawn from a $t$-distribution with $M-1$ degrees of freedom \cite{raghunathan2003multiple}:
\begin{equation}
p\left(\Delta \mid \mu_\Delta, \sigma_\Delta^2\right) \sim t_{M-1}\left(\mu_\Delta, \left(1 + 1/M\right) \sigma_\Delta^2\right),
\label{equation31}
\end{equation}
where the $\sigma_\Delta^2/M$ term in the variance is necessary as an adjustment for there being a finite number of syntheses; as $M \rightarrow \infty$, the variance tends to $\sigma_\Delta^2$.

By performing a large number of simulations, one is estimating the posterior distribution in Equation \ref{equation25} by approximating the integral of the posterior in Equation \ref{equation31} with respect to the posteriors in Equations \ref{equation29} and \ref{equation30} \cite{raghunathan2003multiple}. Hence, the resulting draws of $\Delta$ are samples from the posterior distribution $p\left (\Delta \mid \bm{y}^{\bm{*}}, \bm{t^*} \right)$ in Equation \ref{equation25}. One can take the expectation over the posterior draws to produce a point estimate $\hat{\Delta}$ of the marginal treatment effect in the target distribution of covariates. A point estimate of its variance $\hat{V}\left (\hat{\Delta}\right)$ can be directly computed from the draws of the posterior density. Uncertainty measures such as 95\% interval estimates can be calculated from the corresponding empirical quantiles.

The posterior distributions in Equations \ref{equation29}, \ref{equation30} and \ref{equation31} have been derived under certain normality assumptions, which are adequate for reasonably large sample sizes, where the relevant sample sizes are both the size $N$ of the index study and the size $N^*$ of the synthetic datasets. 


\paragraph{Pooling via combining rules}

A simple alternative to direct Monte Carlo simulation is to use a basic approximation to the posterior density in Equation \ref{equation25}, such that the sampling distribution in Equation \ref{equation31} is normal as opposed to a $t$-distribution. The posterior mean is the average of the treatment effect point estimates across the $M$ syntheses. A combining rule for the variance arises from using $b - \bar{v}$ to estimate $\sigma^2_\Delta$, which is equivalent to setting $\sigma^2_\Delta$ at its approximate posterior mean in Equation \ref{equation30} \cite{si2011comparison}. Again, the $b/M$ term is necessary as an adjustment for there being a finite number of syntheses. 

Consequently, point estimates for the marginal treatment effect in the target covariate distribution and its variance can be derived using the following plug-in estimators:
\begin{equation}
\hat{\Delta}=\bar{\delta},
\label{equation32}
\end{equation}
\begin{equation}
\hat{V}\left(\hat{\Delta} \right)  =\left(1+1/M\right)b - \bar{v}.
\label{equation33}
\end{equation}
The combining rules are slightly different to Rubin's variance estimator in multiple imputation (in Equation \ref{equation33}, $\bar{v}$ is subtracted instead of added) \cite{rubin2004multiple}. Interval estimates can be approximated using a normal distribution, e.g.~for 95\% interval estimates, taking $\pm 1.96$ times the square root of the variance computed in Equation \ref{equation33} \cite{raghunathan2003multiple}. A more conservative, heavier-tailed, $t$-distribution with $\nu_f = \left(M-1 \right) \left(1+\bar{v}/\left(\left(1+1/M\right)b\right)\right)^2$ degrees of freedom has also been proposed, as normal distributions may produce excessively narrow intervals and undercoverage when $M$ is more modest \cite{reiter2002satisfying}. Note that the combining rules in Equations \ref{equation32} and \ref{equation33} are only appropriate for reasonably large $M$. The choice of $M$ is now discussed.

\subsection*{Number of synthetic datasets}

In standard multiple imputation, it is not uncommon to release as little as five imputed datasets \cite{rubin2004multiple}. However, MIM is likely to require a larger value of $M$ because it imputes all of the outcomes in the syntheses, as opposed to a relatively small proportion of missing values. Adopting terminology from the missing data literature, the ``fraction of missing information'' in MIM is 1, because the original dataset used to fit the first-stage regression is different than the augmented target dataset used to fit the second-stage regression. 

In the statistical disclosure limitation literature, a common choice for the number of syntheses is $M=100$ \cite{reiter2002satisfying}. We encourage setting $M$ as large as possible, in order to minimize Monte Carlo error and thereby maximize precision and efficiency. A sensible strategy is to increase the number of syntheses until repeated analyses across different random seeds give similar results, within a specified degree of accuracy. Assuming MCMC simulation is used in the synthesis stage, the value of $M$ is likely to be a fraction of the total number of iterations or posterior samples required for convergence. As computation time is driven by the synthesis stage, increasing $M$ provides more precise and efficient estimation \cite{reiter2002satisfying, reiter2003inference} at little additional cost in the analysis stage. 

An inconvenience of the expressions in Equation \ref{equation30} and Equation \ref{equation33} is that these may produce negative variances. When the posterior in Equation \ref{equation30} generates a negative value of $\sigma^2_\Delta$, i.e., when $\frac{\left(M-1\right)b}{\chi^*} < v$ (where $\chi^*$ is the draw from the posterior in Equation \ref{equation30}), the variance of the posterior distribution in Equation \ref{equation31} is negative. Similarly, Equation \ref{equation33} produces a negative variance when $(1+1/M)b < \bar{v}$. This is because the formulations have been derived using method-of-moments approximations, where estimates are not necessarily constrained to fall in the parameter space. Negative variances are unlikely to occur if $M$ and the size of the synthetic datasets are relatively large. This is due to lower variability in $\sigma^2_\Delta$ and $\hat{V}\left(\hat{\Delta}\right)$ \cite{reiter2005releasing}: $\bar{v}$ decreases with larger syntheses and $b$ is less variable with larger $M$ \cite{reiter2002satisfying}. 

\section*{Simulation study}

\subsection*{Aims}

The objectives of the simulation study are to provide proof-of-principle for MIM and to benchmark its statistical performance against that of the standard implementation of parametric model-based standardization (parametric G-computation), which uses maximum-likelihood estimation with non-parametric bootstrapping for inference \cite{remiro2022parametric, snowden2011implementation, wang2017g, daniel2021making, campbell2023standardization}. The simulation study investigates a setting in which the index study is a perfectly-executed two-arm RCT. This will be standardized to produce a marginal treatment effect in an external target covariate distribution. 

Methods will be evaluated according to the following finite-sample (frequentist) characteristics \cite{morris2019using}: (1) unbiasedness; (2) precision; (3) efficiency; and (4) coverage of interval estimates. The chosen performance metrics assess these criteria specifically. The ADEMP (Aims, Data-generating mechanisms, Estimands, Methods, Performance measures) structure by Morris et al. \cite{morris2019using} is used to describe the simulation study design. Example \texttt{R} code implementing the methods on a simulated example is provided in Additional file 1. All simulations and analyses have been performed using \texttt{R} software version 4.1.1 \cite{team2013r}.\footnote{The files required to run the simulations are available at \url{http://github.com/remiroazocar/MIM}.}

\subsection*{Data-generating mechanisms}

The data-generating mechanisms are inspired by those presented by Phillippo et al. \cite{phillippo2020assessing}. We consider binary outcomes using the log odds ratio as the measure of effect. An index RCT investigates the efficacy of an active treatment (coded as treatment 1) versus a control (coded as treatment 0). Outcome $y_n$ for subject $n$ in the index RCT is simulated from a Bernoulli distribution, with event probabilities $\theta_n = p\left(y_n \mid \bm{x}_n, t_n \right)$ given covariates $\bm{x}_n$ and treatment $t_n$ generated using a logistic model:  
\begin{align*}
y_n &\sim \textnormal{Bernoulli}(\theta_n),    \\
\theta_n &= 
\logit^{-1} \left[
\beta_0 + \beta_{1, 1} x_{n,1}  + \beta_{1, 2} x_{n, 2} +
\left(  \beta_t + \beta_{2, 1} x_{n,1} + \beta_{2,2} x_{n, 2}
\right) 1\left(t_n = 1\right) \right].
\end{align*}
\normalsize

Two correlated continuous covariates, $x_{n, 1}$ and $x_{n, 2}$ are simulated per subject $n$ by drawing from a multivariate Gaussian copula with pre-specified means and standard deviations for the marginal distributions, and a pre-specified covariance matrix. The first covariate follows the marginal distribution $x_{n,1} \sim \textnormal{N}(1, 0.5^2)$, and the second covariate follows the marginal distribution $x_{n,2} \sim \textnormal{N}(0.5, 0.2^2)$. There is some positive correlation between the two covariates, with pairwise correlation coefficients set to 0.15. Both covariates are prognostic of the outcome in the control group at the individual level. Due to the presence of treatment-covariate interactions, both covariates are also (conditional) effect measure modifiers (i.e., predictive of treatment effect heterogeneity) at the individual level on the (log) odds ratio scale. The covariates also modify marginal treatment effects at the population level on the (log) odds ratio scale. 

We set the intercept to $\beta_0 = -0.5$, coefficients for the main covariate-outcome associations to $\beta_{1,k}=2\sigma_{k}$ for covariate $k$, where $\sigma_{k}$ is the standard deviation of the sampling distribution of covariate $k$, and coefficients for the interaction terms to $\beta_{2,k}=\sigma_{k}$. The treatment coefficient (i.e.~the conditional log odds ratio for the active intervention versus control at baseline, when the covariate values are zero) is set to $\beta_{t}=-1.5$. That is, if the binary outcome represents the occurrence of an adverse event, the active treatment would be more efficacious than the control. The covariates could represent influential prognostic and effect-modifying comorbidities that are associated with greater odds of the adverse event and lower efficacy of active treatment versus control at the individual level on the (log) odds ratio scale. 

The simulation study adopts a factorial arrangement using three index trial sample sizes times two levels of overlap between the index trial and the target covariate distributions. This results in a total of six simulation scenarios. The settings are defined by varying the following parameter values: 

\begin{itemize}
    \item Sample sizes of $N \in \{ 500, 1000, 2000\}$ for the index RCT, with a 1:1 active treatment vs.~control allocation ratio.
    \item The level of (deterministic) overlap between the index RCT and the target covariate distribution: limited overlap (50\% of the index study population lies outside of the target population) and full overlap (the index study population is entirely contained within the target population) \cite{phillippo2020assessing}.
\end{itemize}    
Following Phillippo et al. \cite{phillippo2020assessing}, the target covariate distribution is set to achieve the required level of overlap by using a proxy parameter $\kappa$ ($\kappa=0.5$ corresponds to 50\% overlap and $\kappa=1$ corresponds to full overlap). Then, each covariate $k$ in the target follows the marginal distribution $x^{*}_{n,k} \sim \textnormal{N} \left(m^*_k, {\sigma^{*}_k}^{2}\right)$, with $m^*_k = m_k(1.1 + (1-\kappa)^2)$ and $\sigma^*_k = 0.75\sigma_k$, where $m_k$ is the mean of the sampling distribution of covariate $k$ in the index RCT. The target joint covariate distribution is a multivariate Gaussian copula with the pairwise correlation coefficients set to 0.15. $N^{tar}=2000$ subject profiles are simulated for the target covariate dataset. Individual-level outcomes in the target, under the treatments being investigated in the index study, are assumed unavailable and not simulated. 

\subsection*{Estimands}

The target estimand is the true marginal log odds ratio for active treatment versus control in the target covariate distribution. This may vary across the settings of the simulation study because, by design, changing the level of (deterministic) overlap changes the target covariate distribution, and the true marginal log odds ratio depends on the covariate distribution.  

For each scenario, true values of the marginal estimand are determined by simulating a cohort of 2,000,000 subjects, a number sufficiently large to minimize sampling variability, using the target covariate distributions in the simulation study. Hypothetical subject-level binary outcomes under active treatment and control are simulated for the cohort according to the true outcome-generating mechanism. The true marginal log odds ratio is computed by averaging the simulated unit-level outcomes under each treatment and contrasting the marginal outcome expectations on the log odds ratio scale. A simulation-based approach is necessary to compute the true marginal estimands due to the non-collapsibility of the (log) odds ratio \cite{austin2007performance, austin2008performance, remiro2022purely}.

For $\kappa=0.5$ (limited overlap), the true marginal outcome probabilities for active treatment and control in the target population are 0.60 and 0.75, respectively, resulting in a true marginal log odds ratio of -0.68. For $\kappa=1$ (full overlap), the true marginal outcome probabilities for active treatment and control in the target population are 0.50 and 0.69, resulting in a true marginal log odds ratio of -0.81.\footnote{In contrast, the true average conditional log odds ratio for active treatment versus control, at the covariate means of the target population, is given by the weighted average: $\beta_t + \beta_{2, 1} m^*_1 + \beta_{2,2} m^*_2 = -2.5 + 0.5 \times (1.1 + (1-\kappa)^2) + 0.2 \times 0.5 \times (1.1 + (1-\kappa)^2)$. This is equal to -0.69 for $\kappa=0.5$ and to -0.84 for $\kappa=1$.} 

\subsection*{Methods}

Each simulated dataset is analyzed using: (1) the standard implementation of parametric model-based standardization (parametric G-computation) \cite{remiro2022parametric, snowden2011implementation, wang2017g, daniel2021making, campbell2023standardization}; and (2) parametric model-based standardization using MIM. 

\subsubsection*{Standard model-based standardization}

Among the subjects in the index RCT, outcomes are regressed on baseline covariates and treatment using a logistic model. Maximum-likelihood is used to estimate the conditional outcome model, which is correctly specified. Outcome predictions under each treatment are made by applying the fitted regression to the full target covariate dataset. The marginal log odds ratio is derived by: (1) averaging the predicted conditional outcome means by treatment group over the target covariate dataset; (2) transforming the resulting marginal outcome means to the log odds ratio scale; and (3) producing a contrast for active treatment versus control in such scale \cite{remiro2022parametric, daniel2021making, campbell2023standardization}. For inference, the index RCT is resampled via the ordinary non-parametric bootstrap with replacement, using 1,000 resamples (the target covariates are assumed fixed). The average marginal log odds ratio and its standard error are computed as the mean and the standard deviation, respectively, across the resamples. Confidence intervals are computed using the ``percentile'' method; 95\% interval estimates are derived from the 2.5th and the 97.5th percentiles across the resamples.  

\subsubsection*{Multiple imputation marginalization}

In the synthesis stage, the first-stage multivariable logistic regression is correctly specified and is estimated using MCMC sampling. This is implemented using the \texttt{R} package \texttt{rstanarm} \cite{goodrich2020rstanarm}, an appendage to \texttt{rstan} \cite{stanrstan}. We adopt the default normally-distributed ``weakly informative'' priors for the logistic regression coefficients \cite{goodrich2020rstanarm}. Predicted outcomes are drawn from their posterior predictive distribution, given the augmented target dataset. We run two Markov chains with 4,000 iterations per chain, with 2,000 ``burn-in'' iterations that are not used for posterior inference. The MCMC chains are thinned every 4 iterations to use a total of $M=(2000\times 2)/4=1000$ syntheses of size $N^*=2 \times N^{tar}=4000$ in the analysis stage. The second-stage regressions are simple logistic regressions of predicted outcomes on treatment that are fitted to each synthesis using maximum-likelihood estimation. Their point estimates and variances are pooled using the combining rules in Equation \ref{equation32} and Equation \ref{equation33}. Wald-type 95\% confidence intervals are estimated using $t$-distributions with $\nu_f = \left(M-1 \right) \left(1+\bar{v}/\left(\left(1+1/M\right)b\right)\right)^2$ degrees of freedom. Variance estimates are never negative under any simulation scenario. In a test simulation scenario ($\kappa=0.5$ and $N=1000$), the selected value of $M=1000$ is high enough, so that the Monte Carlo error is adequate with respect to the uncertainty in the estimator. Upon inspection, marginal log odds ratio estimates across different random seeds are approximately within 0.01. 

\subsection*{Performance measures}

We simulate 1,000 datasets per scenario. For each scenario and methodology, the following performance measures are computed over the 1,000 simulated datasets: (1) bias; (2) empirical standard error (ESE); (3) mean square error (MSE); and (4) empirical coverage rate of the 95\% interval estimates. These criteria are explicitly defined by Morris et al. \cite{morris2019using} and specifically address the aims of the simulation study. The ESE evaluates precision (aim 2) and the MSE measures overall efficiency (aim 3), accounting for bias (aim 1) and precision (aim 2). 

To quantify the simulation uncertainty, Monte Carlo standard errors (MCSEs) over the data replicates, as defined by Morris et al. \cite{morris2019using}, will be reported for each performance metric. Based on the scenario inducing the highest long-run variability ($\kappa=0.5$ and $N=500$), the MCSE of the bias of the methods is at most 0.015 under 1,000 simulations per scenario, and the MCSE of the coverage (based on an empirical coverage percentage of $95\%$) is $\left(\sqrt{(95 \times 5)/1000}\right)\%=0.69\%$, with the worst-case being $1.58\%$ under $50\%$ coverage. Such levels of simulation uncertainty are considered sufficiently precise, and 1,000 simulations per scenario are deemed appropriate. 

\section*{Results}

Performance metrics for the six simulation scenarios are reported in Figure \ref{fig2}. The limited overlap settings ($\kappa=0.5$) are displayed at the top (in ascending order of index trial sample size, from top to bottom), followed by the full overlap settings ($\kappa=1$) at the bottom. For each simulation scenario, there is a ridgeline plot depicting the spread of point estimates for the marginal log odds ratio across the 1,000 simulation runs. The dashed red lines indicate the true estimands. At the right of each ridgeline plot, a summary tabulation exhibits empirical quantities used to measure the statistical performance of each method, with MCSEs presented in parentheses alongside the corresponding performance metrics.

In the full overlap scenarios, absolute bias is similarly low for MIM and the standard version of model-based standardization. In the limited overlap scenarios, the bias of both standardization methods has slightly higher magnitude. There seems to be a minimal increase in bias as the number of subjects in the index trial decreases. Bias is more marked in the scenarios with $N=500$ (-0.019 and -0.013 for MIM and the standard approach, respectively, in the limited overlap setting, and -0.017 and -0.017, respectively, in the full overlap setting). This is likely due to the small-sample bias inherent in logistic regression \cite{nemes2009bias}. 

As expected, precision is lost as the index trial sample size and the level of covariate overlap are reduced. With $\kappa=0.5$, there exists a subpopulation within the target population that does not overlap with the index trial population. Therefore, inferences in a subsample of the target covariate dataset will rely on extrapolation of the conditional outcome model. With poorer covariate overlap, further extrapolation is required, thereby incurring a sharper loss of precision. Precision is very similar for both standardization methods, as ESEs are virtually equal in all simulation scenarios for both. Similarly, efficiency is virtually identical for both standardization methods. As per the ESE, MSE values also increase as the number of subjects in the index trial and the level of overlap decrease. Because bias is almost negligible across the simulation scenarios, efficiency is driven more by precision than by bias.  

From a frequentist viewpoint, the empirical coverage rate should be equal to the nominal coverage rate to obtain appropriate type I error rates for null hypothesis testing. Namely, 95\% interval estimates should include the true marginal log odds ratio 95\% of the time. Theoretically, due to our use of 1,000 Monte Carlo simulations per scenario, the empirical coverage rate is statistically significantly different to the desired 0.95 if it is less than 0.9365 or more than 0.9635. For both standardization methods, empirical coverage rates only fall outside these boundaries, marginally --- 0.934 for MIM and 0.935 for the standard approach -- in the scenario with $N=500$ and full overlap. This suggests that uncertainty quantification by the standardization methods is adequate. 

\section*{Discussion}

Despite measuring statistical performance in terms of frequentist finite-sample properties, MIM offers performance comparable to that of the standard version of model-based standardization. Both approaches provide appropriate inference with a correctly specified parametric conditional outcome model. The simulation study demonstrates proof-of-principle for the standardization methods, but only considers a simple best-case scenario with correct model specification and two continuous covariates. It does not investigate how robust the methods are to failures in assumptions. 

Parametric outcome models impose strong functional form assumptions; for example, that effects are linear and additive on some transformation of the conditional outcome expectation. Such modeling assumptions may not be plausible where there are a large number of covariates and complex non-linear relationships between them. To provide some protection against model misspecification bias, one may consider using flexible data-adaptive estimators, e.g.~non-parametric or machine learning techniques, for the conditional outcome model (the first-stage regression in MIM). While such approaches make weaker modeling assumptions, they may still be subject to larger-than-desirable bias in finite samples and are constrained by limited theoretical justification for valid statistical inference \cite{naimi2023challenges}. 

In practice, the use of MIM is appealing for several reasons. Firstly, as illustrated in Additional file 1, MIM can readily handle missingness in the patient-level data for the comparative index study. Missing outcomes, and potentially covariate and treatment values, could be imputed in each MCMC iteration of the synthesis stage, naturally accounting for the uncertainty in the missing data of the index study. 

Secondly, the Bayesian first-stage regression model can incorporate both hard external evidence (e.g.~the results of a meta-analysis) and soft external evidence (e.g.~expert knowledge) to construct informative prior distributions for the model coefficients. When external data cannot be leveraged, ``weakly informative'' contextual information can be used to construct skeptical or regularization prior distributions. Through shrinkage, such priors can improve efficiency with respect to maximum-likelihood estimators in certain scenarios \cite{keil2018bayesian}. 

Thirdly, a Bayesian formulation for the first-stage regression offers additional flexibility to address other issues, such as measurement error in the patient-level data of the index trial \cite{keil2014autism}. Bayesian model averaging can be used to capture structural or model uncertainty \cite{madigan1994model}. When one is unsure about which baseline covariates are (conditional) effect measure modifiers, one can allow interactions to be ``half in, half out'' by specifying skeptical prior distributions for the candidate product term coefficients in Equation \ref{equation8} \cite{dixon1991bayesian, spiegelhalter1994bayesian, simon1997bayesian}.\footnote{In the words of Simon and Freedman \cite{simon1997bayesian}, this ``encourages the quantification of prior belief about the size of interactions that may exist. Rather than forcing the investigator to adopt one of two extreme positions regarding interactions, it provides for the specification of intermediate positions.''} 

In this article, we have used multiple imputation to perform model-based standardization over a target empirical covariate distribution, assumed to belong to participants that are external to the index study. In this scenario, one is reliant on correct specification of the outcome model (the first-stage regression in MIM) for unbiased estimation in the target. One is also reliant on covariates being consistently defined across data sources and on complete information on influential covariates being available both for the index study and for the target. In practice, this is a key challenge \cite{stuart2015assessing, stuart2017generalizing}, which could be addressed through the development of core patient characteristic sets that define clinically important covariates to be measured and reported among specific therapeutic areas \cite{vuong2023development}. 

In the absence of overlap between the covariate distributions in the index study and the external target, e.g.~when the index study covariate distribution lies outside the target covariate distribution, one must consider the plausibility of the outcome model extrapolation. Sensitivity analyses using alternative model specifications may be warranted to explore the dependence of inferences on the selected adjustment model. Recently, several authors have proposed sensitivity analyses that are applicable where potential effect measure modifiers are measured only in the index trial but not in the target dataset \cite{nguyen2017sensitivity, dahabreh2019extending}. These techniques could be applied in conjunction with MIM. 

While we have used multiple imputation to standardize over an external covariate distribution, it is also possible to standardize over the empirical covariate distribution of the index study, as illustrated in Additional file 1. This involves less stringent assumptions and avoids model-based extrapolation into an external data source. Such approach allows for the estimation of covariate-adjusted marginal treatment effects within individual comparative studies, adjusting for covariate imbalances between treatment arms.

A limitation of this article is the lack of a real case study demonstrating the application of the new methodology. While proof-of-principle for MIM has been provided through simulation studies,
the method should be applied to a real example in order to influence applied practice. This is a key
priority for future research.





\begin{backmatter}

\section*{Declarations}

\vspace{0.3cm}

\section*{Acknowledgments}

Not applicable.

\section*{Funding}

AH is funded by Canada Research Chair in Statistical Trial Design; Natural Sciences and Engineering Research Council of Canada (award No. RGPIN-2021-03366).




\section*{Author contributions}

ARA, AH and GB conceived the research idea, developed the methodology, performed the analyses, prepared the figures, and wrote and reviewed the manuscript.

\section*{Abbreviations}

ADEMP -- Aims, Data-generating mechanisms, Estimands, Methods, Performance measures \\
DAG -- Directed acyclic graph \\
ESE -- Empirical standard error \\
IPD -- Individual patient data \\
HTA -- Health technology assessment \\
MCMC -- Markov chain Monte Carlo \\
MCSE -- Monte Carlo standard error \\
MIM -- Multiple imputation marginalization \\
MSE -- Mean square error \\
RCT -- Randomized controlled trial \\

\section*{Availability of data and materials}

The files required to generate the data, run the simulations, and reproduce the results are available at \url{http://github.com/remiroazocar/MIM}.

\section*{Ethics approval and consent to participate}

Not applicable.

\section*{Competing interests}
  
We declare that the authors have no competing interests as defined by BMC, or other interests that might be perceived to influence the results and/or discussion reported in this paper.

\section*{Consent for publication}

Not applicable. 
  

\bibliographystyle{bmc-mathphys} 
\bibliography{bmc_article}      




\clearpage

\section*{Figures}

\begin{figure}[!htb]
  \includegraphics[width=0.9\textwidth]{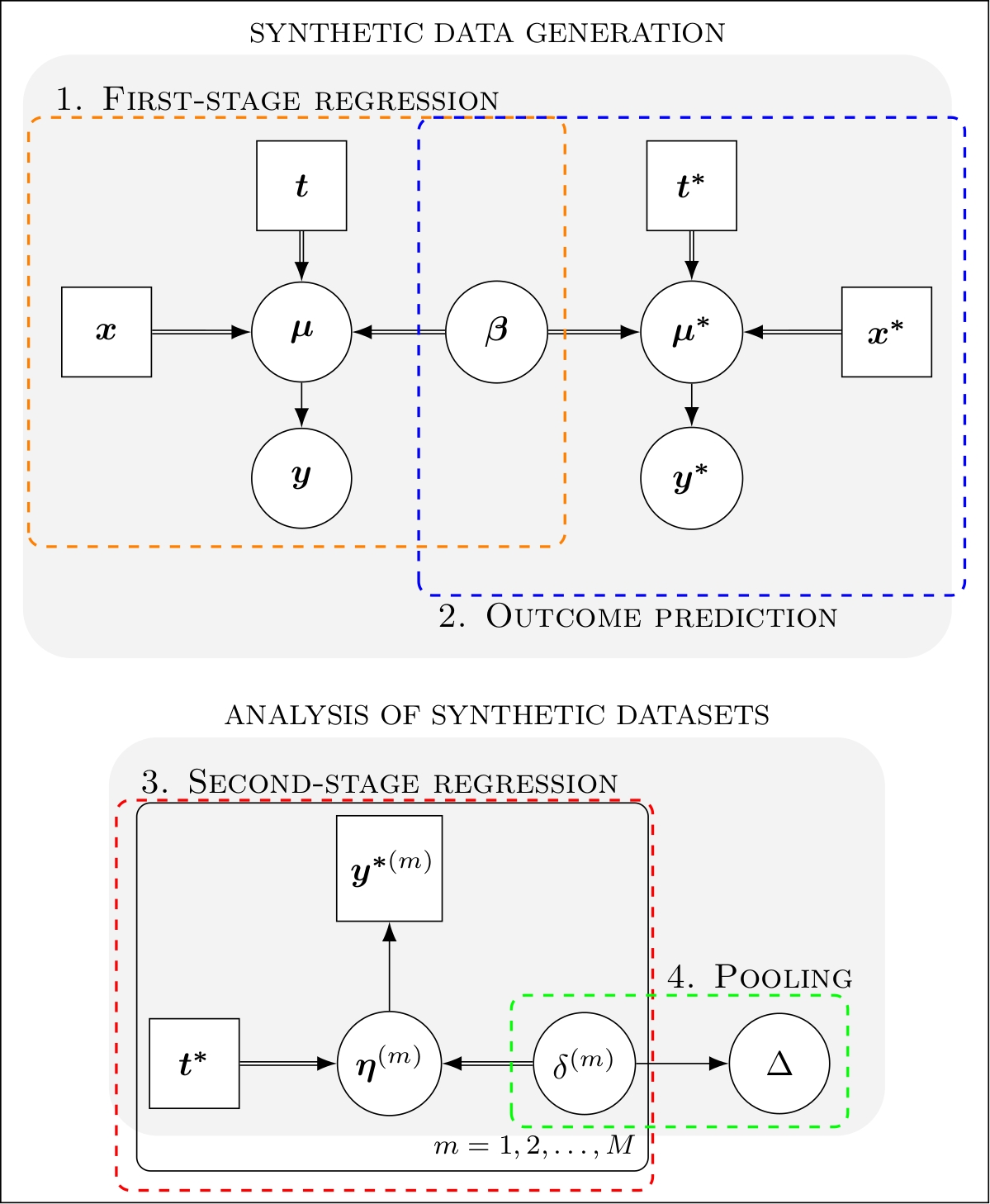}
  \caption{\csentence{Multiple imputation marginalization (MIM).}
A Bayesian directed acyclic graph representing MIM and its two main stages: (1) synthetic data generation; and (2) the analysis of synthetic datasets. Square nodes represent constant variables, circular nodes indicate stochastic variables, single arrows denote stochastic dependence, double arrows indicate deterministic relationships and the plate notation indicates repeated analyses.}
     \label{fig1}
\end{figure}

\begin{figure}[!htb]
  \includegraphics[width=0.98\textwidth]{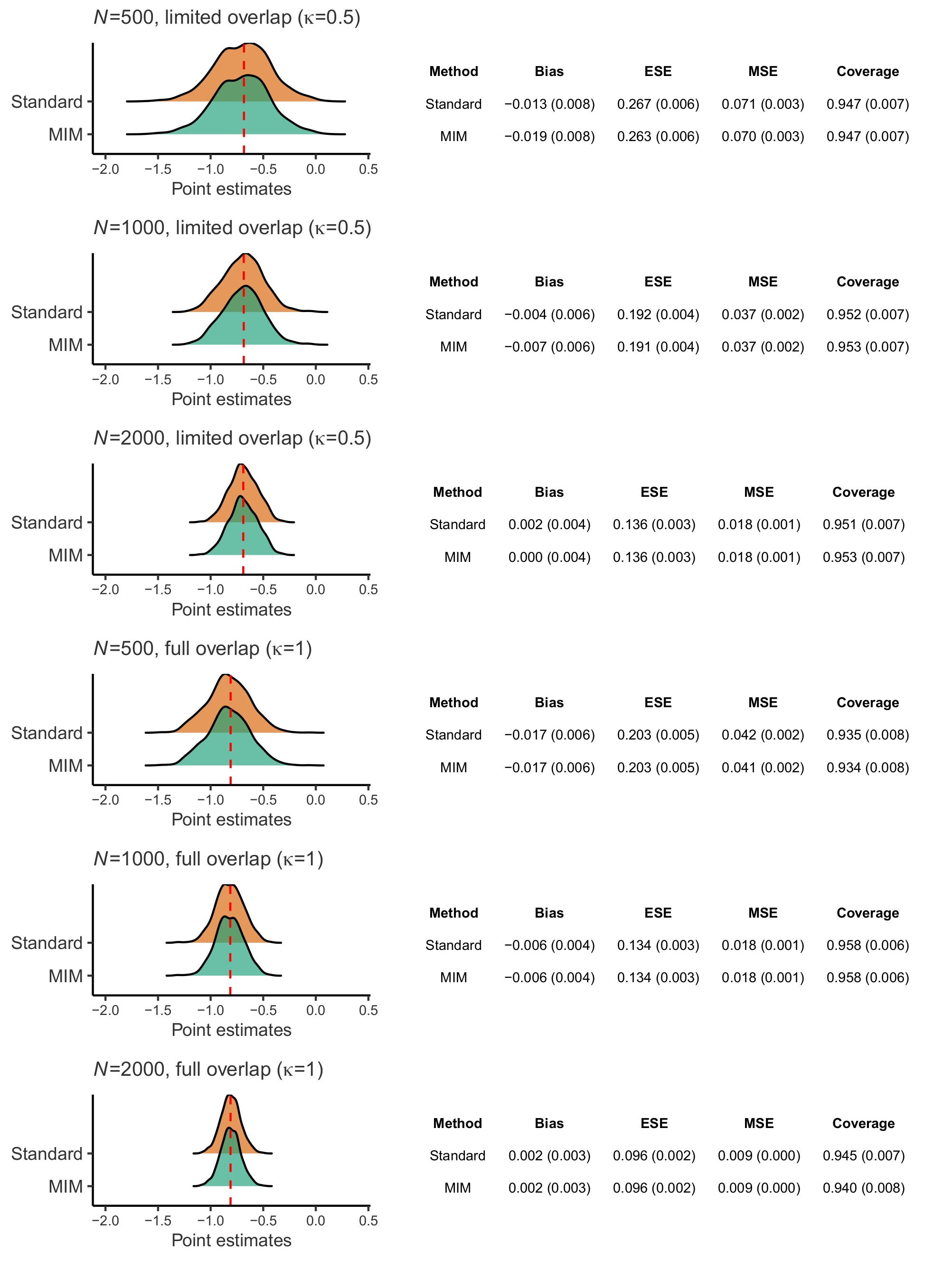}
  \caption{\csentence{Simulation study results.} The distribution of treatment effect point estimates over the simulation runs and the empirical quantities used to measure the statistical performance of standard model-based standardization (``Standard'') and multiple imputation marginalization (``MIM'') are visualized for the six scenarios. The dashed red lines in the ridgeline plots to the left indicate the true estimands.}
     \label{fig2}
\end{figure}

\clearpage


\section*{Tables}

\begin{table}[!htb]
\caption{An example of the structure of the $m$-th synthetic dataset, created in the data synthesis stage of MIM. In this example, $N=7$ ($N^*=14$) and $K=3$. Prior to imputing the missing outcomes, a copy of the  original target covariate dataset has been assigned the treatment value zero and been vertically concatenated to the original $\bm{x}^{tar}$, assigned the treatment value one.}
\centering
\begin{tabular}{lcccccr}
\hline
\multicolumn{3}{l}{Covariates ($\bm{x^*}$)} &  Treatment ($\bm{t^*}$) & Outcome $\left (\bm{y}^{\bm{*}(m)}\right )$\\
\hline
$x_{1,1}^{tar}$ & $x_{1,2}^{tar}$ & $x_{1,3}^{tar}$ & 1 & $y^{*(m)}_1$ \\
$x_{2,1}^{tar}$ & $x_{2,2}^{tar}$ & $x_{2,3}^{tar}$ & 1 & $y^{*(m)}_2$ \\
$x_{3,1}^{tar}$ & $x_{3,2}^{tar}$ & $x_{3,3}^{tar}$ & 1 & $y^{*(m)}_3$ \\
$x_{4,1}^{tar}$ & $x_{4,2}^{tar}$ & $x_{4,3}^{tar}$ & 1 & $y^{*(m)}_4$ \\
$x_{5,1}^{tar}$ & $x_{5,2}^{tar}$ & $x_{5,3}^{tar}$ & 1 & $y^{*(m)}_5$ \\
$x_{6,1}^{tar}$ & $x_{6,2}^{tar}$ & $x_{6,3}^{tar}$ & 1 & $y^{*(m)}_6$ \\
$x_{7,1}^{tar}$ & $x_{7,2}^{tar}$ & $x_{7,3}^{tar}$ & 1 & $y^{*(m)}_7$ \\
\hline
$x_{1,1}^{tar}$ & $x_{1,2}^{tar}$ & $x_{1,3}^{tar}$ & 0 & $y^{*(m)}_8$ \\
$x_{2,1}^{tar}$ & $x_{2,2}^{tar}$ & $x_{2,3}^{tar}$ & 0 & $y^{*(m)}_9$ \\
$x_{3,1}^{tar}$ & $x_{3,2}^{tar}$ & $x_{3,3}^{tar}$ & 0 &  $y^{*(m)}_{10}$ \\
$x_{4,1}^{tar}$ & $x_{4,2}^{tar}$ & $x_{4,3}^{tar}$ & 0 &  $y^{*(m)}_{11}$ \\
$x_{5,1}^{tar}$ & $x_{5,2}^{tar}$ & $x_{5,3}^{tar}$ & 0 &  $y^{*(m)}_{12}$ \\
$x_{6,1}^{tar}$ & $x_{6,2}^{tar}$ & $x_{6,3}^{tar}$ & 0 &  $y^{*(m)}_{13}$ \\
$x_{7,1}^{tar}$ & $x_{7,2}^{tar}$ & $x_{7,3}^{tar}$ & 0 &  $y^{*(m)}_{14}$ \\
\hline
\end{tabular}
\label{tab1}
\end{table}




\end{backmatter}

\end{document}